\documentclass[conference]{IEEEtran}
\IEEEoverridecommandlockouts
\usepackage{cite}
\usepackage{amsmath,amssymb,amsfonts}
\usepackage{algorithmic}
\usepackage{graphicx}
\usepackage{textcomp}
\usepackage{xcolor}

\def\BibTeX{{\rm B\kern-.05em{\sc i\kern-.025em b}\kern-.08em
		T\kern-.1667em\lower.7ex\hbox{E}\kern-.125emX}}

\begin{document}
	
	\title{Cryptography: Post-Quantum versus Classical
	}
	
	\author{\IEEEauthorblockN{Abhinav Awasthi}
		\IEEEauthorblockA{
			\textit{Pranveer Singh Institute of Technology}\\
			Kanpur, Uttar Pradesh \\
			2k22.csai.2211953@gmail.com}
			
	\and
	
	\IEEEauthorblockN{Atul Chaturvedi}
	\IEEEauthorblockA{
		\textit{Pranveer Singh Institute of Technology}\\
		Kanpur, Uttar Pradesh \\
		atul.chaturvedi@psit.ac.in}
	}
	
	\maketitle
	
	\begin{abstract}
		The advantages of post-quantum cryptography over classical cryptography are covered in this survey. We address several post-quantum cryptography techniques. We conclude that the deployment of quantum-safe cryptographic systems is anticipated to be the future of secure communication, and that the development of post-quantum cryptography is essential to guarantee the security of sensitive information in the post-quantum era.
	\end{abstract}
	
	
	\section{Introduction}
Understanding the implications of cutting-edge technologies like quantum computing is crucial because cyber security is more crucial than ever. In addition to influencing how businesses can prepare and safeguard themselves with quantum-resistant encryption, quantum computing has the potential to significantly impact the future of cyber security. The development of quantum computing will have a significant impact on cyber security, particularly on current encryption standards. The ability of quantum computing to break various forms of public-key cryptography, such as RSA encryption, which is utilized by numerous industries globally, is one of the technology's most contentious uses. Researchers are actively advancing the field of post-quantum cryptography to bolster the security of digital communication in anticipation of quantum computing threats. This burgeoning area explores alternative mathematical foundations, such as lattice-based or hash-based cryptography, aiming to devise encryption methods resilient to quantum algorithms that could compromise current cryptographic standards. The urgency arises from the potential of algorithms like Shor's to break widely-used protocols like RSA and ECC. As interdisciplinary collaborations progress, the goal is to establish new cryptographic standards that ensure the enduring security of information in the post-quantum era [1 - 12].
	
	\section{Cryptography}
	A secure communication method in situations involving malevolent attackers and unapproved users is cryptography. There are two steps in the cryptography process:
	\begin{enumerate}
		\item \textbf{Encryption:} The process of employing keys (an extra data string) to transform crucial data or a message (known as plaintext, represented by M) into an unintelligible encrypted form (known as ciphertext, represented by C).It is carried out by the sender.
		
		\item \textbf{Decryption:} The process of translating plaintext M from ciphertext C using the same or a different key in order to decode the message.It is carried out by the recipient.
		
	\end{enumerate}
	
	\subsection{Classical Cryptography}
	The security of classical cryptography, which has its roots in mathematics, depends on how hard it is to compute mathematical problems like factorizing big numbers \cite{b13}. Traditional cryptography employs two categories of methods:
	
	\begin{enumerate}
		\item \textbf{Symmetric Cryptography:} Data encryption and decryption require the sharing of a single private key, which can only be used by the authorized sender and recipient.
		
		\item \textbf{Asymmetric Cryptography:} The public key and private key are the two keys used in encryption and decryption. Data is encrypted by the sender using a public key, and it is decrypted by the recipient using a private key. With this method, the key distribution issue that plagues symmetric cryptography is resolved.
		
	\end{enumerate}
	
	Since the encryption process followed by decryption process recovers the original plaintext, the following must hold true to make our protocol correct:
	$D(E(M)) = M$
	
	\subsection{Key pillars of classical cryptography}
	
	\begin{enumerate}
		\item \textbf{Confidentiality:} Ensuring that information is kept secret and protected from unauthorized access. Classical cryptographic techniques, such as encryption algorithms, are employed to maintain the confidentiality of data.
		
		\item \textbf{Integrity:} Verifying the integrity of the information to ensure that it has not been altered or tampered with during transmission or storage. Techniques like message authentication codes (MACs) and hash functions contribute to maintaining data integrity.
		
		\item \textbf{Authentication:} Confirming the identity of communication parties, ensuring that the sender or recipient is who they claim to be. Classical cryptographic methods, including digital signatures and certificates, are utilized for authentication purposes.
		
		\item \textbf{Non-repudiation:} Preventing individuals from denying their involvement in a communication or transaction. Digital signatures and other cryptographic techniques help establish non-repudiation by providing proof of origin or receipt.
	\end{enumerate}
	
	\subsection{Cryptography Algorithms}
	 An algorithm is a mathematical formula that is used in cryptography to jumble and make unintelligible plain text. Its purpose is to prevent unwanted access to sensitive data. These algorithms are frequently utilized in many different applications, including online transactions, data storage, and secure communication. For the purpose of maintaining the integrity and confidentiality of the encrypted data, they rely on intricate mathematical operations and keys. 
	 
	 \textbf{Modular Arithmetic:} Modular arithmetic is a system of arithmetic for integers where numbers "wrap around" when reaching a certain value, called the modulus. When we divide two integers we will have an equation that looks like the following
	 $A/B = Q \text{ remainder} R$ where A is the dividend, B is the divisor, Q is the quotient and R is the remainder. We can extract the remainder using an operator called the modulo operator (abbreviated as mod).
	 $A \mod B = R \text{ (A modulo B is equal to R) }$B is referred to as the modulus. For example,
	 $13 \mod 5 = 3$. Modular Addition is the sum of two integers X and Y followed by a modulo operation with a modulus B, $(X + Y) \mod B$.

	 \textbf{One-Time Pad:} One-Time Pad is an example of Symmetric Cryptography.
	 One-Time Pad is an encryption technique that cannot be decrypted. Each bit / character of the plaintext is encrypted using the corresponding bit / character from the paired key (one-time pad) using modular addition \cite{b14}.
	 
	\textbf{Shift Cipher:} A Shift Cipher (for example, the Caesar Cipher) shifts every letter by the same shift (some number between 1 and 26). If someone called Alice was to encrypt her name, it would result in one of 26 possible encryptions. This is a small number of possibilities, and all can be checked using brute force attack (checking all possibilities for a solution) \cite{b15}.
	
	\textbf{Shannon's Perfect Secrecy:} In 1946, Claude Shannon defined the idea of perfect secrecy, which means that the ciphertext C, gives no information about the plaintext (except the maximum possible length of the message). Given a truly random key, a ciphertext can be translated into any plaintext, and all are equally possible \cite{b16}.
	
	\textbf{RSA Encryption Scheme: }RSA encryption scheme is an example of Asymmetric Cryptography. A pair of keys (public key and private key) is used for encryption and decryption. A sender uses a public key to encrypt the data and receiver can decrypt the data using his or her private key. This technique overcomes the problem of key distribution as in symmetric cryptography. RSA is named on the initials of its creators: Rivest, Shamir, and Adleman \cite{b17}.
	
	\textbf{Diffie-Hellman Key Exchange:} The Diffie-Hellman key exchange, developed by Whitfield Diffie and Martin Hellman in 1976, revolutionized cryptography by enabling two parties to establish a shared secret key over an insecure public channel without ever directly exchanging the key itself \cite{b17}. 
	
	\section{Post Quantum Cryptography}
	Quantum computers have the potential to break many of the existing public-key cryptographic systems that depend on hard mathematical problems. Post-quantum cryptography (PQC) is a proactive initiative to develop cryptographic algorithms that remain impregnable even in the face of quantum adversaries. Unlike its conventional counterpart, PQC doesn't rely on the unproven "hardness" of specific mathematical problems.
	
	\subsection{Need for Post Quantum Cryptography}
	Our reliance on classical public cryptography for securing data is facing a significant threat due to the advent of quantum computing and quantum algorithms. Classical encryption relies on the assumption that certain mathematical problems are practically impossible for classical computers to solve, given their computational limitations and the absence of efficient algorithms. For instance, RSA encryption hinges on the challenge of factoring large integers into two prime numbers. While multiplying these primes to obtain a large integer is easy, the reverse – factoring the large integer – poses a formidable challenge. This mathematical complexity forms the foundation of widely used public cryptography protocols like RSA and ECC. However, the landscape is changing with the emergence of quantum computing, which has the potential to break these classical encryption methods, posing a substantial threat to the security of our data and information.
	
	\subsection{Post Quantum Cryptographic Algorithms}
	PQC algorithms encompass various approaches:
	
	\begin{itemize}
		\item \textbf{Lattice-Based Cryptography:} Lattice-based cryptography is predicated on the difficulty of specific mathematical lattice problems, which are geometric structures made up of repeated point patterns in several dimensions. The difficulty of solving specific lattice problems, like the Learning With Errors (LWE) problem, is the foundation for the security of lattice-based schemes \cite{b18}.
		
		\item \textbf{Code-Based Cryptography:} Error-correcting codes serve as the basis for cryptographic schemes in code-based cryptography. The foundation of security lies in the difficulty of deciphering a linear code without being aware of the error-correction feature. One popular illustration of a code-based cryptographic algorithm is the McEliece cryptosystem \cite{b18}.
		
		\item \textbf{Hash-Based Cryptography:} The idea of hash-based digital signatures and hash functions are the foundation of hash-based cryptography. Because the security is based on hash function characteristics, it is impervious to attacks even in the event that quantum computers are present. In hash-based schemes, the Merkle-Damgård construction is frequently employed \cite{b19}.
		
		\item \textbf{Multivariate Polynomial Cryptography:} Systems of multivariate polynomials over finite fields are used in multivariate polynomial cryptography. The difficulty of solving nonlinear equation systems forms the basis of the security. Multivariate polynomial cryptographic algorithms include the Rainbow and Hidden Field Equations (HFE) schemes \cite{b20}.
		
		\item \textbf{Quantum Superposition and Quantum Entanglement:} These are quantum mechanical ideas rather than particular cryptographic algorithms. Post-quantum cryptographic algorithms have been impacted by the ability of quantum computers to exist in multiple states concurrently, which is made possible by quantum superposition. Correlated states between particles are made possible by quantum entanglement, and these states can be investigated in cryptographic protocols.
		
	\end{itemize}
	
	\section*{Conclusion}
In conclusion, the urgency surrounding the exploration of post-quantum cryptography is paramount, as the rapid advancement of quantum computers poses an imminent and grave threat to the very fabric of current cryptographic security. The escalating vulnerability of traditional public-key systems to quantum attacks underscores an immediate need for the development of robust cryptographic algorithms. Across the globe, researchers are fervently delving into numerous post-quantum cryptography techniques, with the National Institute of Standards and Technology (NIST) leading the charge in identifying and standardizing these crucial methods. The relentless collaboration among international cryptography experts not only underscores the severity of the situation but also reflects a shared commitment to fortify digital security in the face of an ever-evolving technological landscape. As NIST advances in its standardization efforts, the imperative to implement quantum-safe cryptography systems becomes increasingly dire. In a world where digital communication intricately weaves the fabric of modern society, the prospect of quantum threats looms ominously, necessitating the swift adoption of recommended standards for a semblance of reassurance. Post-quantum cryptography not only responds to an immediate peril but also marks a pivotal juncture in the ongoing and dynamic interplay between technological progress and the safeguarding of cryptographic security. To ensure a resilient foundation for the long-term security of data in this ever-changing digital environment, stakeholders across industries must confront the gravity of the situation head-on, displaying unwavering vigilance, flexibility, and proactive adherence to recommended standards. The specter of quantum threats demands nothing less than a concerted and immediate response to fortify the digital realm against an impending and formidable challenge.

	\vspace{12pt}
	\color{red}
\end{document}